\documentclass[%
reprint,
superscriptaddress,
 amsmath,amssymb,
 aps,
pra,
]{revtex4-2}

\usepackage{graphicx}
\usepackage{dcolumn}
\usepackage{bm}
\usepackage{comment}

\usepackage[utf8]{inputenc}
\usepackage{times}
\usepackage{float}
\usepackage{xcolor}

\begin{document}

\preprint{APS/123-QED}

\title{Monochromatization of Electron Beams with Spatially and Temporally Modulated Optical Fields}

\author{Neli Laštovičková Streshkova}
\email[]{neli.streshkova@matfyz.cuni.cz}
\author{Petr Koutenský}
\affiliation{Department of Chemical
Physics and Optics, Faculty of Mathematics and Physics, Charles University, Ke Karlovu 3, Prague CZ-12116, Czech Republic.}%
\author{Tomáš Novotný}
\affiliation{Department of Condensed Matter Physics, Faculty of Mathematics and Physics, Charles University, Ke Karlovu 5, Prague CZ-12116, Czech Republic.}%
\author{Martin Kozák}
\email[]{m.kozak@matfyz.cuni.cz}
\affiliation{Department of Chemical
Physics and Optics, Faculty of Mathematics and Physics, Charles University, Ke Karlovu 3, Prague CZ-12116, Czech Republic.}%


\date{\today}

\begin{abstract}
Inelastic interaction between coherent light with constant frequency and free electrons enables periodic phase modulation of electron wave packets leading to periodic side-bands in the electron energy spectra. In this Letter we propose a generalization of the interaction by considering linearly chirped electron wave packets interacting with chirped optical fields. We theoretically demonstrate that when matching the chirp parameters of the electron and light waves, the interaction leads to partial monochromatization of the electron spectra in one of the energy side-bands. Depending on the coherence time of the electrons, the electron spectrum may be narrowed down by a factor of 5-times with 26\% of the electron distribution in the monochromatized energy band. This approach will improve the spectral resolution and reduce color aberrations in ultrafast imaging experiments with free electrons.

\end{abstract}

\maketitle

In recent years, coherent control of the electron wave function utilizing optical fields has been extensively explored. Electromagnetic field of light can serve as a tool for accelerating electrons \cite{Breuer2013,Peralta2013,Chlouba2023,Shiloh2021,Kozak2022,Black2019}, generating attosecond electron pulses \cite{Schonenberger2019,Vanacore2018,Feist2015,Tsarev2023,Kozak2018,Kozak2019,Tsarev2023,Black2019,Priebe2017}, tailoring the electron quantum states \cite{Kozak2021,Reinhardt2020,Priebe2017,Kozak2018_Phys,Shiloh2022,Piazza2015,Dahan2021,Dahan2020_Nature,Echternkamp2016,Andrea2023,Feist2020,Chirita2022} or for applications in time-resolved electron diffraction, interferometry and spectroscopy experiments \cite{Gracia2010_Rev,Baum2007,Baum2009,Kozak2017,Feist2017,Harvey2020}. 

To achieve efficient interaction between the electron wave function and the light field in table-top set-ups it is necessary to overcome the energy-momentum mismatch that is present in vacuum. This can be done using evanescent fields of waveguide structures \cite{Henke2021,Wang2020,Kfir2020}, optical near-fields of nanostructures \cite{Feist2015,Gracia2016,Talebi2020} or semi-infinite optical fields \cite{Vanacore2018,Andrea2023,Feist2020}. Alternatively, efficient control can be also achieved in vacuum without the need for a special nanostructure or mirror with the use of optical beat waves synchronized with the free electron wave packet \cite{Baum2007,Kozak2018,Kozak2019,Kozak2018_Phys,Tsarev2023}.

A general interaction of free electrons with quasi-monochromatic optical fields leads to a periodic phase modulation of the electron wave function and consequently to the generation of equidistant side-bands in the electron energy spectra. The peaks are separated by the energy $\hbar\omega$, which corresponds to the energy of emitted or absorbed photons \cite{Feist2015,Park2010,Tsarev2023}.

This Letter proposes a generalization of the quantum coherent control of electron wave packets using modulating light field with time-dependent frequency enabling monochromatization of a significant portion of the electron distribution. Electrons emitted from the electron source in TEM or SEM have a finite coherence length that corresponds to the spectral width of about 0.5-1 eV \cite{BAUM2013,Zewail_4D_microscopy}. The finite spectral width leads to elongation of the electron pulse during its dispersive propagation between the photocathode and the sample. The electron pulse thus acquires an energy chirp that can be described as time-correlated energy shift \cite{BAUM2013,Zewail_4D_microscopy}. Interestingly, the instantaneous energy width of a chirped electron wave packet is smaller than its full energy width. This allows for spectral squeezing of the electron pulse by compensating the electron wave packet chirp in the energy-time domain using its inelastic interaction with chirped optical modulating field. The development of low-loss monochromatization techniques for pulsed electrons is important for improving the spectral resolution, for example in applications such as quantum sensing with free electrons \cite{Karnieli2023}. 

\begin{figure*}
\includegraphics[width=\textwidth]{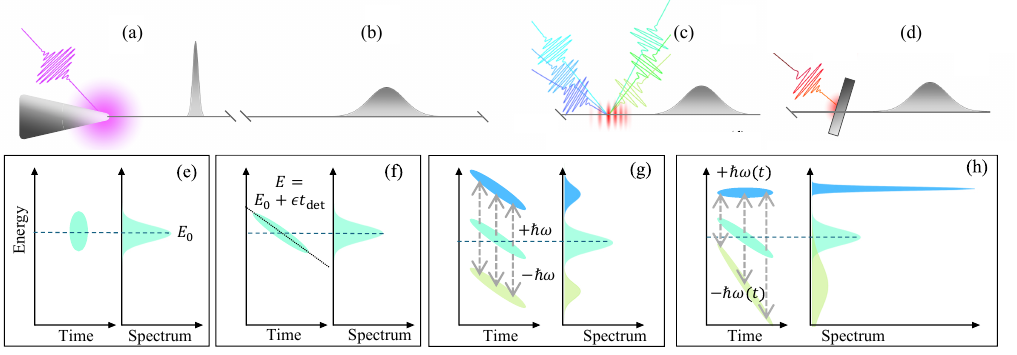}
\caption{Schematic illustration of the proposed method of electron beam monochromatization. (a) Photo-emission of the electron pulse from the source triggered by an UV optical pulse. The temporal duration of the electron pulse immediately after photoemission corresponds to the UV pulse duration. (b) The electron pulse elongates in time due to dispersive propagation in vacuum. (c) The elongated electron pulse interacts with the ponderomotive potential of a chirped optical beat wave formed by two spatio-temporally chirped optical pulses. (d) Alternative interaction with semi-infinite optical fields excited by a single chirped optical pulse at a semi-transparent membrane. (e) Electron pulse immediately after photoemission and (f) after dispersive propagation.  (g) Interaction with a quasi-monochromatic optical field and (h) with a chirped optical pulse, where the chirp parameter matches the electron pulse chirp.
}
\label{fig:01}
\end{figure*}

In the proposed scheme we consider an electron pulse photo-emitted from a cathode of an electron gun by a femtosecond UV pulse (see Fig. \ref{fig:01}(a)). The energy-time (momentum-coordinate) representation of the electron pulse immediately after the emission is shown in Fig. \ref{fig:01}(e), where the vertical major axis of the ellipse corresponds to the pulse's spectral width, and the horizontal minor axis corresponds to the pulse's temporal duration. The electron wave packet then propagates dispersively which leads to temporal elongation of the pulse (Fig. \ref{fig:01}(b)) and the ellipse representing the electron pulse in energy-time space deforms accordingly (Fig. \ref{fig:01}(f)). For a narrow relative spectral width of the electron pulse the group velocity dispersion can be considered to be linear. After the deformation the initial spectral width is conserved as well as the phase space volume. The major axis of the ellipse is parallel to the the line $E(t)=E_{0}+\epsilon t$, where $E_{0}$ is the initial central energy of the wave packet and $\epsilon < 0$ denotes the chirp coefficient and $E_{0}$ is the center energy of the electron distribution.

Coherent interaction between the chirped electron pulse and quasi-monochromatic optical field with time independent photon energy $\hbar\omega$ yields multiple population side-bands shifted in energy by $\pm\hbar\omega$, which are characterized by the same linear chirp as the initial electron wave packet \cite{Kirchner2013}. The generation of the first side-bands corresponding to emission or absorption of one photon is schematically shown in (Fig. \ref{fig:01}.(g)). The spectrum of the wave packet after the interaction contains the side-bands with the same spectral width as the original electron pulse.

By using the time-dependent frequency of optical modulating fields, we can more intricately manipulate the resulting electron spectra. We propose a generalization of the inelastic electron scattering scheme either by using the ponderomotive potential of two chirped optical laser beams (Fig. \ref{fig:01}(c)) on which we focus in this Letter. The principle of energy side-band generation with time-variant photon energy is general and any configuration allowing longitudinal phase modulation of the electrons by optical fields can be used. For example, we illustrate a semi-infinite light field scheme with a single chirped optical pulse (Fig. \ref{fig:01}(d)); similarly, a photon-induced near-field electron microscopy scheme could facilitate the interaction.

The concept of electron monochromatization is illustrated in Fig. \ref{fig:01}(h). The time dependent frequency of the modulating light field results in  a time dependent instantaneous energy shift of the side-bands by $\hbar\omega(t)$. This allows to transfer a part of the chirped electron pulse to a nearly horizontal distribution in one of the energy side-bands while conserving the instantaneous energy width of the electron distribution in the chirped pulse. Because of the fact that the instantaneous energy width can be significantly narrower than the overall spectral width of the electron pulse immediately after photoemission we reach monochromatization of the electrons in the side-band (blue spectrum in Fig. \ref{fig:01}(h)).

We note that the scheme proposed here is conceptually different from recently demonstrated monochromatization of electron pulses by THz fields \cite{Kaminer2023}, where the effect is based on classical chirp compensation by time-dependent force acting on a classical electron distribution.

We describe the interaction of the free electron wave packet and the chirped optical fields within the semi classical framework. The free electron Hamiltonian in vacuum takes the form
\begin{equation}
\label{eqn:01}
    \hat{H} = \frac{1}{2m_{e}}\left( \hat{\mathbf{p}}+e\mathbf{A}\right)^{2}=\hat{H}_{0}+\hat{H}_{\text{int}},
\end{equation}
where $m_{e}$ is the electron mass, $\hat{\textbf{p}}$ is the electron momentum operator, $e>0$ is the elementary charge and $\textbf{A}$ is the classical vector potential of the electric field. The free-space evolution of the electron wave packet is given by $\hat{H}_0=\hat{\textbf{p}}^2/2m_{e}$ and is treated by introducing the Dirac picture, with the density matrix defined as $\hat{\rho} = e^{i\hat{H}_{0}t / \hbar}\hat{\rho}_{S} e^{-i\hat{H}_{0}t/\hbar}$, where $\hat{\rho}_{S}$ is the density matrix in the Schr\"{o}dinger picture. The interaction with the optical field is given by $\hat{H}_{\text{int}}=e(\textbf{A}\cdot\hat{\textbf{p}}+\hat{\textbf{p}}\cdot\textbf{A})/2m_{e}+e^{2}\textbf{A}^{2}/2m_{e}$. 

The state of the electron wave packet is described in the framework of the Wigner function in the momentum-coordinate phase space, accounting for its partially incoherent nature. Calculations are conveniently done in the electron rest frame, where $z = z_{\text{lab}}-v_{0}t$ (we assume sub-relativistic group velocity $v_{0}$ of the electrons). Considering the non-recoil approximation (negligible change of electron velocity during the interaction) and the nearly linear dispersion of electrons accelerated to high kinetic energy we obtain the Wigner representation in energy-time phase space by linearly transforming the axes $\Delta E = E-E_{0}=pv_{0}$, $t=-z/v_{0}$. The energy-time representation is a snapshot of the electron wave packet at a constant coordinate $z_{\text{lab}}$ in the laboratory frame, which we set to zero without loss of generality. The electron spectral density is calculated by integrating the Wigner function along the time $t$ axis. For a detailed derivation of the Wigner function and the electron-optical field interaction; see the Supplemental Material (SM) \cite{SM}.

Prior to interaction with the optical field, the chirped electron wave packet can be described analytically by the Wigner function
\begin{equation}
\label{eqn:02}
W_{\text{in}}=\exp\left[-
\begin{pmatrix}
    z, & p
\end{pmatrix}
\cdot
\Sigma^{-1}
\cdot
\begin{pmatrix}
z\\
p
\end{pmatrix}
\right],
\end{equation}
with the correlation matrix
\begin{equation}
\label{eqn:03}
    \Sigma^{-1} = 
    \begin{pmatrix}
    \frac{1}{v_{0}^{2}(\sigma_{c}^{2}+\sigma_{s}^{2})} & -\frac{\sigma_{p}\sigma_{c}}{\hbar^(\sigma_{c}^{2}+\sigma_{s}^{2})}\\ \\
    -\frac{\sigma_{p}\sigma_{c}}{\hbar^(\sigma_{c}^{2}+\sigma_{s}^{2})} & \frac{ v_{0}^{2}\sigma_{c}^{2}(\sigma_{c}^{2}+\sigma_{p}^{2}+\sigma_{s}^{2})}{\hbar^{2}(\sigma_{c}^{2}+\sigma_{s}^{2})}
    \end{pmatrix}.
\end{equation}
The electron wave packet is represented by a tilted ellipse as in Fig. \ref{fig:01}(f), which is defined by three temporal parameters - the coherence time $\sigma_{c}$ (inversely proportional to the initial total energy width), the incoherent smearing time $\sigma_{s}$ corresponding to the electron emission time uncertainty (proportional to the duration of the UV photoemission laser pulse) and the pulse elongation $\sigma_{p}$ due to the chirp acquired during propagation. To model the Wigner function of the wave packet before acquiring the chirp through dispersive propagation we set $\sigma_{p}=0$.

By performing Weyl transformation (\cite{SM}, Eq. (S25)) on the Wigner function we obtain the analytical formula for the density matrix in the coordinate representation $\langle z|\hat{\rho}_{\text{in}}|z'\rangle$. The interaction of the density matrix with the optical field is calculated as
\begin{equation}
\label{eqn:04}
    \langle z|\hat{\rho}_{\text{out}}| z'\rangle=\hat{U}(z) \langle z|\hat{\rho}_{\text{in}}| z'\rangle\hat{U}^{\dagger}(z').
\end{equation}
The evolution is modelled within the non-recoil approximation as a phase modulation induced by the interaction Hamiltonian \cite{Gracia2021}
\begin{equation}
\label{eqn:05}
    \hat{U}(z)=\exp\left[-\frac{i}{\hbar}\int^{\infty}_{-\infty} H_{\text{int}}(z,t')\,\mathrm{d}t'\right],
\end{equation}
where the electron rest frame coordinate $z$ is treated as a parameter. In vacuum due to the fast oscillations of the optical field the terms in $\hat{H}_{\text{int}}$ containing $\hat{\textbf{p}}$ do not contribute to the phase modulation, as the integral in Eq. (\ref{eqn:05}) averages out to zero. The nonzero contribution comes from the term $e^{2}\textbf{A}^{2}/2m_{e}$ describing the ponderomotive potential generated by the two optical pulses. We note that the derivation is analogous in alternative schemes (photon-induced near-field electron microscopy, semi-infinite light fields), where the coupling is enabled through the $\hat{\textbf{p}}\cdot \textbf{A}$ and $\textbf{A} \cdot  \hat{\textbf{p}}$ terms and the ponderomotive potential can be neglected.

The total vector potential is given by the sum of the vector potentials corresponding to the optical beams $\textbf{A}=\textbf{A}_{1}+\textbf{A}_{2}$. We consider the vector potentials to be harmonic functions oscillating at central frequencies $\omega_{0,1}$, $\omega_{0,2}$ with slowly varying envelopes and small chirps. Then the interaction Hamiltonian can be rewritten using the electrical fields $\textbf{E}_{1}$, $\textbf{E}_{2}$ (for details see SM \cite{SM})
\begin{equation}
\label{eqn:06}
    H_{\text{int}} = 
\sum_{i,j=1}^{2}\frac{e^{2}}{2m\omega_{0,i}\omega_{0,j}}\mathbf{E}_{i}^{}\cdot\mathbf{E}^{*}_{j}.
\end{equation}

We model the chirped optical pulses as plane waves with Gaussian envelopes in space and time polarized along the $x$ axis:
\begin{equation}
\label{eqn:07}
\begin{split}
    &\mathbf{E}_{i}^{}(\mathbf{r},t) =\mathbf{e}_{x} \frac{E_{0,i}}{2}e^{-\frac{\xi_{i}^{2}}{\sigma_{\mathrm{opt},i}^{2}}}e^{-i\omega_{0,i}\xi_{i}-i a_{i}\xi_{i}^{2}(\alpha_{i})}
    \times\\
    &\exp\Bigg[-\frac{x^{2}+\left(y\cos\alpha_{i}-z_{\text{lab}}\sin\alpha_{i}\right)^{2}}{w_{i}^{2}}\Bigg] 
    + c.c.,
\end{split}
\end{equation}
where $E_{0,i}$ is the amplitude, $\xi_{i} = t-(y \sin\alpha_{i} + z_{\text{lab}}\cos\alpha_{i})/c$ is the wave argument of the beam propagating along the angle $\alpha_{i}$ in the $yz_{\text{lab}}$ plane with respect to the $z_{\text{lab}}$ axis, $c$ is the speed of light, $ \sigma_{\text{opt,i}}$ is the temporal half-width of the electric field, $a_{i}$ is the linear chirp and $w_{i}$ is the radius of the beam. We transform the optical fields into the electron rest frame, plug into Eqs. (\ref{eqn:06}) and (\ref{eqn:05}) and numerically calculate the evolution operator that induces oscillating phase modulation with $z$-dependent local frequency $(\omega_{0,1}-\omega_{0,2})/v_{0}+2(a_{1}-a_{2})z/v_{0}^{2}$ onto the density matrix. Finally, we obtain the Wigner function of the electron wave packet after the inelastic ponderomotive interaction $W_{\text{out}}$ via applying the Wigner transformation (\cite{SM}, Eq. (S26)) numerically on $\langle z|\hat{\rho}_{\text{out}}| z'\rangle$. 

We consider a partially coherent electron wave packet interacting with the ponderomotive potential of a chirped optical beat wave formed by two chirped pulses with different center frequencies. Specifically, we consider energy transitions involving purely longitudinal momentum transfer, generalizing previous studies \cite{Baum2007,Baum2009,Kozak2018,Kozak2019,Kozak2018_Phys,Tsarev2023}. This requires two optical beams with photon energies $\hbar\omega_{1}$, $\hbar\omega_{2}$ and angles of incidence $\alpha_{1}$, $\alpha_{2}$ relative to the electron propagation axis $z$, fulfilling the condition \cite{Kozak2018}
\begin{equation}\label{eqn:08}
    \frac{c}{v_{0}} = \frac{\omega_{1}\cos{\alpha_{1}}-\omega_{2}\cos{\alpha_{2}}}{\omega_{1}-\omega_{2}},
\end{equation}
The condition for zero transverse momentum transfer is \cite{Kozak2018}
\begin{equation}
\label{eqn:09}
    \omega_{1}\sin{\alpha_{1}}-\omega_{2}\sin{\alpha_{2}}=0.
\end{equation}
Ideally, Eqs. (\ref{eqn:08}) and (\ref{eqn:09}) must be satisfied for every position in space and every instance in time. When operating with chirped optical fields, we need to take into account the frequencies changing with the time of arrival of the individual spectral components to the interaction region. The angles need to compensate for that change to ensure the phase matching for a sufficiently long time window, which should cover at least the duration of the electron pulse. Otherwise, if the angles are kept constant, Eqs. (\ref{eqn:08}) and (\ref{eqn:09}) are fulfilled only for the central frequencies and Eq. (\ref{eqn:05}) induces non-zero phase modulation only along a short window around the center of the electron pulse.

We consider the phase-matching only on the $z_{\text{lab}}$ axis ($x$=$y$=0), a valid approximation for an electron beam much narrower than the optical beams. To reach monochromatization of one of the electron spectral side-bands we assume linearly chirped optical fields with  instantaneous frequencies $\omega_{i}(t) = \omega_{0,i}+2a_{i}t$. The electron energy side-bands at each time are shifted in energy-time phase space by $\hbar(\omega_{1}(t)-\omega_{2}(t))$. The central energy of the first side-band of the distribution is then described by $E(t)=E_{0}+\epsilon t+\hbar(\omega_{0,1}-\omega_{0,2})+2\hbar(a_{1}-a_{2})t $. If the terms linear in time cancel out, we achieve a horizontal side-band with constant central energy, requiring the following condition for the chirp parameters of optical and electron beams
\begin{equation}
    \label{eqn:10}
    \epsilon = 2\hbar (a_{2}-a_{1}).
\end{equation}

\begin{figure}
\includegraphics[]{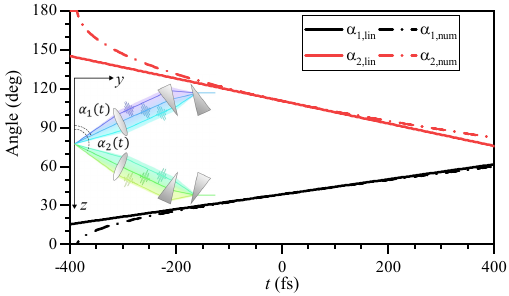}
\caption{Time dependencies of the phase-matching angles $\alpha_{1}$ and $\alpha_{2}$ calculated numerically (dot and dashed lines) and analytically in the linear approximation (solid lines). Time zero is the center of the chirped incoherent electron pulse. The inset shows a possible scheme for experimental implementation.}
\label{fig:02}
\end{figure}

The light fields interacting with the electron wave packet have central angular frequencies $\omega_{0,1}=5.47\,\mathrm{fs}^{-1}$, $\omega_{0,2}=3.65\,\mathrm{fs}^{-1}$ (wavelengths of $343 \,\mathrm{nm}$ and $515\,\mathrm{nm}$ corresponding to the second and third harmonic frequencies of ytterbium-based laser systems) and linear chirp $a_{1,2}= \pm 7.6\cdot10^{-4}\,\mathrm{fs}^{-2}$. The energy-time tilt of the electron probability distribution is $\epsilon = -\frac{0.5\,\mathrm{eV}}{250\,\mathrm{fs}}$. The velocity of the electron wave packet is $v_{0}=0.32c$ corresponding to the kinetic energy of $30\,\mathrm{keV}$. For the given simulation parameters we have calculated numerically the angles required for phase matching $\alpha_{1}(t)$, $\alpha_{2}(t)$ if Eqs. (\ref{eqn:08}) and (\ref{eqn:09}) are fulfilled at each $t$ (see dashed curves in Fig. \ref{fig:02}). When considering realistic means of realization, we can expect approximately linear time dependence of the angles of incidence that correspond to an angularly chirped beam, which can be experimentally generated by placing a pair of prisms and a focusing lens (layout is shown in the inset of Fig. \ref{fig:02}). The time dependence of the angles fulfilling the synchronicity condition can be approximated by linear function for the duration of the interaction (for formula see the Supplemental Material \cite{SM}). 

\begin{figure}[h]
    \centering
    \includegraphics{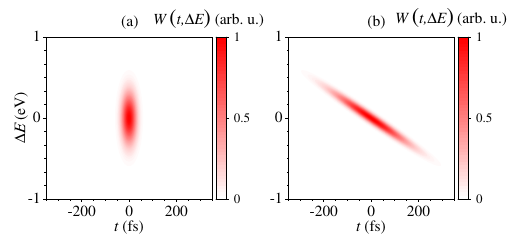}
    \caption{Normalized Wigner function $W(t, \Delta E)$ of the electron beam (a) right after emission from the electron source, (b) after acquiring linear chirp through propagation.}
    \label{fig:03}
\end{figure}

Now we relate the sigma values for the calculations in correspondence to the experimentally obtainable FWHM time duration. The FWHM spectral width of $\Delta E_{\text{FWHM}}=0.5\,\mathrm{eV}$ corresponds to the coherence time $\sigma_{c}=2.19\,\mathrm{fs}$. Considering a statistical mixture from different emission times, the pulse has an overall FWHM duration of $\tau_0=50\,\mathrm{fs}$, with a temporal broadening parameter $\sigma_{s}=29.9\,\mathrm{fs}$. The normalized Wigner function of the electron wave packet after emission from the source is shown in Fig. \ref{fig:03}(a). The elongation in time gives total FWHM duration of the chirped pulse $\tau_{\text{chirped}}=250\,\mathrm{fs}$, which gives $\sigma_{p}=147.13\,\mathrm{fs}$. The Wigner function of the chirped electron pulse after propagation to the site of the interaction with optical fields is shown in Fig. \ref{fig:03}(b).

    The temporal FWHM duration of both light pulses is $\tau_{\text{opt}}=250\,\mathrm{fs}$ and the Gaussian spot radii are $w_{1}=5\,\mathrm{\mu m}$ and $w_{2}=3\,\mathrm{\mu m}$.

\begin{figure}[h]
    \centering
    \includegraphics{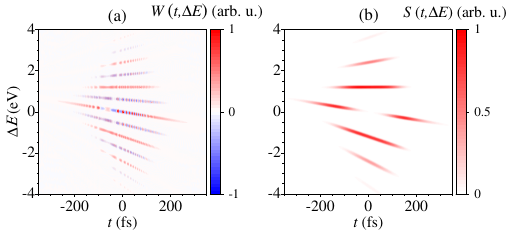}
    \caption{(a) Normalized Wigner function $W(t, \Delta E)$ of the electron beam after interaction with the chirped optical pulses containing rapidly oscillating coherences between the states. (b) Spectrogram $ S(t, \Delta E)$ of the electron beam after interaction with the chirped optical pulses representing the populations of the energetic side-bands.}
    \label{fig:04}
\end{figure}

The Wigner function of the electron wave packet after interaction with the light fields is shown in Fig. \ref{fig:04}(a). Multiple energy side-bands are generated at both higher and lower energies than the initial electron energy $E_0$. The Wigner function of a non-classical state typically has negative values. It can be shown that in the generation of energetic side-bands in the free electron distributions, these features arise from coherences between the individual side-bands (see analytical insight in SM \cite{SM}).

By convolving $W_{\text{out}}$ with a Gaussian kernel, we average over the coherences and obtain a non-negative spectrogram $S = W_{\mathrm{out}} \ast e^{-z^{2}/\Delta ^{2}}$, $\Delta \approx 900\, \mathrm{nm}$. The spectrogram represents the instantaneous populations of the free electron energy states as shown in Fig. \ref{fig:04}(b). Around time $t=0$, where the optical pulse overlap is the largest and the phase-matching is satisfied almost perfectly, we observe a significant fraction of the initial electron population transferred to the side-bands and depletion of the zero loss peak. By compensating the electron chirp, we achieve a narrow horizontal peak in the first energy gain side-band centered at $\Delta E=1.2\,\mathrm{eV}$.

\begin{figure}[h]
    \centering
    \includegraphics{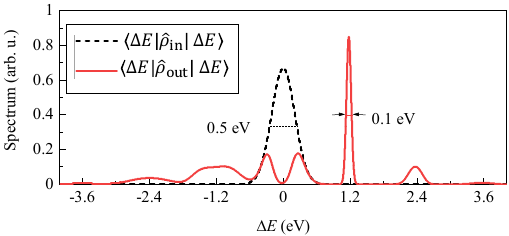}
    \caption{Spectrum of the electron wave packet before $\langle \Delta E|\hat{\rho}_{\text{in}}|\Delta E\rangle$ (black, dashed) and after $\langle \Delta E|\hat{\rho}_{\text{out}}|\Delta E\rangle$ (red, full) monochromatization. The monochromatized peak is squeezed by a factor of 5.}
    \label{fig:05}
\end{figure}

The electron spectra before and after the interaction are plotted in Fig. \ref{fig:05}. The FWHM of the monochromatized peak centerred at $\Delta E=1.2\,\mathrm{eV}$ is $0.1\,\mathrm{eV}$. The factor of the spectral squeezing can be estimated using Eqs. (\ref{eqn:02}) and (\ref{eqn:03}). The coherent interaction transfers parts of the electron distribution between the side-bands vertically, conserving the local vertical width of the original distribution $W_{\text{in}}$. Setting $z=0$ we estimate that the side-peak is squeezed by a factor of $(\sigma_{c}^{2}+\sigma_{s}^{2}+\sigma_{p}^{2})^{\frac{1}{2}}/(\sigma_{c}^{2}+\sigma_{s}^{2})^{\frac{1}{2}}$, which is the ratio of the duration of the pulse with and without chirp.
 
The demonstrated monochromatization by a factor of 5 is achieved with experimentally feasible parameters used in the simulations. It can be improved by increasing the electron pulse chirp, leading to a narrower local energy width of the Wigner function. This can be achieved either by a longer propagation distance or by a weaker extraction field at the photocathode. Additionally, at these parameters, 26\% of the electrons are in the monochromatized peak, significantly reducing current losses compared to conventional monochromators.

There are two main practical limitations of the proposed monochromatization method. As the initial spectral width of the electron wave packet increases, the spectral widths of the interacting photons must also increase. Given that the center photon energy of optical fields is about 2-4 eV, the maximum spectral width of the optical beat wave is limited to a few eV. For broader electron wave packets, the scheme can be modified to monochromatize higher-order energy side-bands, where the optical field chirp is effectively multiplied by the side-band index $N$. The second limitation is that, for broad spectra, both the electron pulse chirp and optical pulse chirp are typically nonlinear. This chirp curvature generates a curved first-order side-band, deteriorating monochromatization performance.

In this Letter, we numerically simulate monochromatization for parameters typical of SEMs with pulsed laser-triggered electron sources. With adjustments to optical field frequencies, the technique is also applicable to electrons accelerated to hundreds of keV, typical for TEMs. 

In summary, we presented a method for monochromatizing chirped electron wave packets through their interaction with generalized optical fields of time-dependent frequency. We detailed the possibility of spectral squeezing via inelastic ponderomotive scattering. However, the principle of electron wave function manipulation is broader and applicable to different phase-matching schemes, such as semi-infinite optical fields. The technique enables quantum coherent manipulation of free electrons with spatially and temporally modulated optical fields, offering new possibilities for tailoring free electron quantum states in both time and energy domains.

\section*{Acknowledgements}

The authors acknowledge funding from the Czech Science Foundation (project 22-13001K), Charles University (SVV-2023-260720, PRIMUS/19/SCI/05, GAUK 90424) and the European Union (ERC, eWaveShaper, 101039339). Views and opinions expressed are however those of the author(s) only and do not necessarily reflect those of the European Union or the European Research Council Executive Agency. Neither the European Union nor the granting authority can be held responsible for them.
This work was supported by TERAFIT project No. CZ.02.01.01/00/22\_008/0004594 funded by OP JAK, call Excellent Research.


\providecommand{\noopsort}[1]{}\providecommand{\singleletter}[1]{#1}%

\end{document}